\documentstyle[aps,epsfig]{revtex}
\begin{document}
\tightenlines
\draft

\newcommand{\be}{\begin{eqnarray}}
\newcommand{\ee}{\end{eqnarray}}
\newcommand{\dia}{\!\!\!\!\!\not\,\,\,}

\twocolumn[\hsize\textwidth\columnwidth\hsize\csname
@twocolumnfalse\endcsname

\title{Boundary and expansion effects on two-pion correlation
       functions in relativistic heavy-ion collisions} 
\author{Alejandro Ayala and Angel S\'anchez}
\address{Instituto de Ciencias Nucleares\\
         Universidad Nacional Aut\'onoma de M\'exico\\
         Apartado Postal 70-543, M\'exico Distrito Federal 04510, M\'exico.}
\maketitle
\begin{abstract}
We examine the effects that a confining boundary together with
hydrodynamical expansion play on two-pion distributions in
relativistic heavy-ion collisions. We show that the effects arise from
the introduction of further correlations due both to collective motion
and the system's finite size. As is well known, the former leads to a
reduction in the apparent source radius with increasing
average pair momentum $K$. However, for small $K$, the presence of the
boundary leads to a decrease of the apparent source radius with
decreasing $K$. These two competing effects produce a maximum for the
effective source radius as a function of $K$.  
\end{abstract}
\vskip2pc]

\section{Introduction}\label{I}
In recent years, much experimental effort has been devoted to the
production of a state of matter under the extreme conditions of
high baryonic densities and/or temperatures in relativistic heavy-ion
collisions. The main drive for such effort is the expectation to
produce the so called quark-gluon plasma (QGP) where the fundamental QCD
degrees of freedom are not confined within a single nucleon but rather
over a larger volume of the order of the dimensions of a
nucleus. While the properties of the QGP have been the subject of
intense theoretical study and debate, much less attention has been
paid to the hadronization process, or to the properties of hadronic
matter at high temperatures and densities. An understanding of these
is needed for a correct interpretation of the signals originating from
the different stages of the collision, both for a clear distinction of
a possible QGP formation but also as an interesting subject of study
on its own.

The most abundantly produced hadrons in relativistic heavy-ion
collisions are pions. Typically, the number of pions produced one unit
around central rapidity in central Au+Au collisions at energies of
order 10$A$ GeV is $dN_\pi /dy\sim 300$\cite{E-802}. Under the
assumption that the transverse dimensions of the system formed at
freeze out are of order of the transverse size of an Au nucleus
and that the typical pion formation time is of order 1 fm, this
multiplicity implies that the average pion separation $d$ at freeze
out in the central rapidity region is of order $d\sim 0.6$ fm which is
less than the average range of the pion strong interaction, $d_s\sim
1.4$ fm.

Some of the possible consequences of this large pion density produced
in relativistic heavy-ion collisions where first studied by
Shuryak~\cite{Shuryak} who coined the term {\it pion liquid} to refer
to the situation where the pion system could not be thought of as
existing as a hadron gas but rather, that its properties resembled
more those of a liquid of quasipions. In
particular, as one of the main characteristics of liquids is the
appearance of a surface tension, such state of hadron matter could
give rise to a confining boundary that acted as a reflecting surface
that could affect the pion distributions just before freeze out. More
recently~\cite{Kostyuk}, it has been proposed that the equation of
state of pion matter could give rise to a phase transition from a gas
phase to a more dense phase as the temperature rises close to the
temperatures expected to be achieved in relativistic heavy-ion
collisions, thus introducing the concept of a {\it hot pion liquid}. 

From a phenomenological point of view and disregarding the details of
the reflection processes (which presumably depend on the energy of the
incident particle), the development of a surface tension can be
incorporated by imposing a sharp boundary for the pion wave functions
to evolve just before freeze out. As a consequence of the finite size
of the system during this stage of its evolution, the energy states
form a discrete set. 

An important difference between statistical systems with and without
a boundary is the different density of states at increasing energies,
being larger in the case of the former, as illustrated in Fig.~1. The
density of states of a finite system approaches that of an unbound
one as the size of the system is increased. The above characteristic
implies that the transverse inclusive spectrum calculated
within a boundary model will exhibit a concave shape at high
transverse momentum and could potentially explain the increase of the
pion transverse distribution~\cite{increase} in this kinematical 
region~\cite{Wong}. The increase of the pion distribution at low
transverse momentum could also be explained within the same context by
considering the finite chemical potential associated with the mean pion
multiplicity per event~\cite{Ayala1}.

More recently, it was realized that an important missing ingredient 
in the description of the transverse pion spectra within a boundary
model was the proper inclusion of hydrodynamical expansion. The
phenomenological description in terms of a bound, expanding pion
system has been named the {\it expanding pion liquid} model and
was developed in Refs.~\cite{Ayala2} and successfully applied to the
description of the experimental mid-rapidity, transverse pion spectra
in central Au+Au collisions at 11.6$A$ GeV/c~\cite{E-802}. A further
natural test ground for the model is the study of multiparticle
correlations, in particular two-pion correlations. 

%%%%%%%%%%%%%%%%%%%%%%%%%%%%%%
\begin{figure}[t] % fig 1
\centerline{\epsfig{file=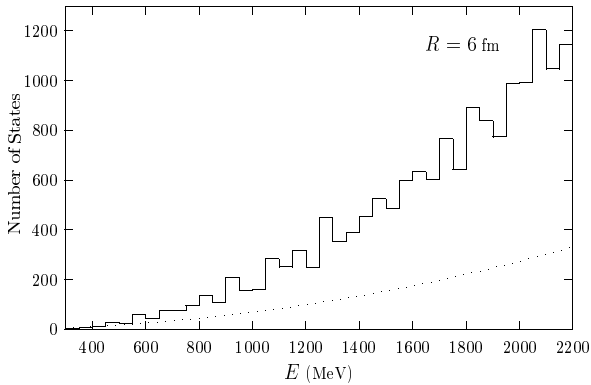,height=2.0in,width=3in}
}
\vspace{0.5cm}
\caption{Number of states for a spherically symmetric system with a
         sharp boundary at $R=6$ fm (solid line) compared to a system
         without boundary (dotted line). Notice how the presence of
         the boundary makes the number of states grow faster at large
         energies.}
\end{figure}
%%%%%%%%%%%%%%%%%%%%%%%%%%%%%

A step in this direction has been taken in Ref.~\cite{Zhang} where the
effects of a pure boundary model ({\it i.e.} without
collective expansion) have been introduced in the description of the
two-pion correlation function. In this work, we incorporate also the
effects of hydrodynamical expansion in the calculation of the two-pion
correlation function. 

It is well known that in the study of
two-particle correlation functions, the effective size of a
system without a boundary decreases as the average pair momentum is
increased~\cite{Pratt} when considering the effects of hydrodynamical
expansion. Physically, this effect is due to the fact that as the
average pair momentum increases, the particles in the pair are more
likely to be emitted from points close in space. This can also be
regarded as the introduction of a further correlation in phase space
for the emitted particles that effectively reduces the size of the
emission region~\cite{Zajc}. On the other hand, 
as emphasized in Ref.~\cite{Zhang}, for emission volumes of order
of a few average pair wavelengths it is imperative to consider
a full quantum density matrix in the description of particle
distributions since as the average pair momentum decreases, the
correlation region as a function of relative momentum increases thus
effectively reducing the apparent size of the system. As we will show
in this work, these two competing effects produce a maximum in the
effective size of the system at a finite value of the average pair momentum.

This paper is organized as follows: Section~\ref{II} is devoted to a
recollection of the description of a bound, expanding pion system at
freeze out. In Sec.~\ref{III}, we write the two-pion correlation
function in terms of the discrete set of eigenfunctions for this kind
of systems. In Sec.~\ref{IV} we perform a systematic analysis of
the two-pion correlation function in terms of the different parameters
involved. We pay particular attention to the behavior of the effective
system's radius as a function of the average pair momentum comparing
the results to what would be expected in the case of an expanding and
unbound system and a pure boundary model. Finally we conclude and
discuss our results in Sec.~\ref{V}.

\section{The expanding pion liquid}\label{II}
When the system of pions can be considered as confined, its wave
functions satisfy a given condition a the boundary. In order to
compare the results with the observed particle distributions, the
shape of the assumed confining volume could play an important role. It 
has long been known that the particle momentum distributions are somewhat
forward-backward peaked, particularly at energies of the
SPS~\cite{KLM}, even in the case of central collisions. Nevertheless,
for the sake of simplicity and concreteness, here we will assume that
the confining volume is spherical. Comparison of the model with data
will become better in the central rapidity region where an asymmetry
between longitudinal and transverse expansion is less important than
in the fragmentation region. Let us emphasize that an spherically
symmetric model is not essential to the basic physics discussed here
and can thus be relaxed at the expense of additional computing
time. This could however be necessary when comparing to data away from
mid-rapidity. 

To incorporate the effects of an hydrodynamical flow, we notice that
this ordered motion can be represented by a four-velocity field 
$u^{\mu}=\gamma (r)[1,{\mathbf v}(r)]$, where $\gamma$ is the
Lorentz-gamma factor and ${\mathbf v}(r)$ is the velocity vector. This
field represents a redistribution of momentum in each of the fluid
cells, as viewed from a given reference frame (the center of mass in
this case), becoming centered around the momentum associated with the
velocity of the fluid element. This behavior can be described by the
substitution $p^\mu\rightarrow p^\mu-mu^\mu$, where $m$ is the pion
mass. The term $mu^\mu$ represents the collective momentum of the
given pion fluid element.

The eigenfunctions of the confined, expanding system of pions are thus
obtained as the solutions to the equation
\be
   \Big\{ -\left[i\frac{\partial}{\partial t}-m\gamma (r)\right]^2
   &+& \Big[-i\nabla-m\gamma (r){\mathbf v}(r)\Big]^2
   \nonumber\\
   &+& m^2\Big\}\psi ({\mathbf r},t)=0
   \label{wavefunc}
\ee
where we look for the stationary states subject to the boundary condition
\be
   \psi(|{\mathbf r}|=R,t)=0\, .
   \label{bondcond}
\ee
We consider a parametrization of the velocity vector ${\mathbf v}(r)$
that scales with the distance from the center of the fireball
\be
   {\mathbf v}(r)=\beta\frac{r}{R}{\mathbf\hat{r}}\, ,
   \label{paramvel}
\ee
where the parameter $0 < \beta <1$ represents the surface fireball
hydrodynamical velocity. Correspondingly, the explicit expression for
$\gamma (r)$ becomes
\be
   \gamma (r)=\frac{1}{\sqrt{1-\beta^2r^2/R^2}}\, .
   \label{gammaexpl}
\ee
Since Eq.~(\ref{wavefunc}), with the gamma factor given by
Eq.~(\ref{gammaexpl}), can only be solved numerically, we resort to
approximate the function $\gamma$ by the first terms of its Taylor
expansion
\be
   \gamma (r) \simeq 1 + \frac{\beta}{2}\frac{r^2}{R^2}\, ,
   \label{gammaapprox}
\ee
which is valid for not too large values of $\beta$. With this
approximation, Eq.~(\ref{wavefunc}) becomes an equation for particles
moving in a spherical harmonic well with rigid boundaries and can be
solved analytically. The stationary states are 
\be
   \psi_{nlm'}({\mathbf r},t)&=&\!\frac{A_{nl}}{\sqrt{2E_{nl}}}e^{-iE_{nl}t+
   im\beta r^2/(2R)}Y_l^{m'}(\hat{\mathbf r})e^{-\alpha_{nl}^2r^2/2}
   r^l\nonumber \\
   &\times&\!_1F_1\left(\frac{(l+3/2)}{2}-
   \frac{\varepsilon_{nl}^2}{4\alpha_{nl}^2}, 
   l+3/2;\alpha_{nl}^2r^2\right)\!\!,\!\!
   \label{solnew}
\ee
where $\,_1F_1$ is a confluent hypergeometric function and $Y_l^{m'}$ is a 
spherical harmonic. The quantities $A_{nl}$ are the normalization constants 
and are found from the condition
\be
   \int d^3r \psi^{\ast}_{nlm'}({\mathbf r},t)
   \frac{\stackrel{\leftrightarrow}{\partial}}{\partial t}
   \psi_{nlm'}({\mathbf r},t) = 1\, .
   \label{norma}
\ee
The parameters $\alpha_{nl}$ and $\varepsilon_{nl}$ are related to the energy 
eigenvalues $E_{nl}$ by
\be
   \alpha^4_{nl}&=&m(E_{nl}-m)\beta^2/R^2\, ,\nonumber\\
   \varepsilon^2_{nl}&=&E_{nl}(E_{nl}-2m)\, ,
   \label{param}
\ee
with $E_{nl}$ given as the solutions to
\be
   \,_1F_1\left(\frac{(l+3/2)}{2}-\frac{\varepsilon_{nl}^2}{4\alpha^2_{nl}},
   l+3/2;\alpha_{nl}^2R^2\right)=0\, .
   \label{encond}
\ee
Equation~(\ref{solnew}), along with the energy eigenvalues and
definitions in Eqs.~(\ref{param}),~(\ref{encond}), constitute the 
set of (properly normalized) eigenfunctions in terms of which the
various multi-particle distributions can be expressed. The system's
finite size and the strength of the collective expansion are given in
terms of the parameters $R$ and $\beta$, respectively. In order to
further proceed, it is necessary to specify the kind of 
ensemble that describes the statistical properties of the pion system. 

\section{Two-pion correlation function}\label{III}
In order to describe the situation where equilibrium has been attained
(which we assume here), the proper statistical distribution for the ensemble
is thermal. For the purposes of this section, we closely follow
Ref.~\cite{Zhang} to where we refer the reader for details. Let
$\lambda$ represent the set of quantum numbers $\{nlm'\}$. The
corresponding occupation number $N_\lambda$ for a given state is written as
\be
   N_\lambda=\frac{1}{\exp (E_\lambda-\mu)/T -1}\, ,
   \label{occupation}
\ee
where $T$ is the system's temperature and $\mu$ the chemical
potential, related to the average total multiplicity $N$ by
\be
   N=\sum_\lambda\frac{1}{\exp (E_\lambda-\mu)/T -1}\, .
   \label{mult}
\ee
Let $\psi_\lambda ({\mathbf p})$ represent the Fourier transformed
wave function for the state with quantum numbers $\lambda$, namely
\be
   \psi_\lambda ({\mathbf p})=\int d^3r\ e^{-i{\mathbf p\cdot r}}
   \psi_\lambda ({\mathbf r})\, .
   \label{FT}
\ee
With the normalization adopted in Eq.~(\ref{solnew}), the one-pion
momentum distribution can be written as  
\be
   P_1({\mathbf p})&\equiv&\frac{d^3N}{d^3p}\nonumber\\
   &=&\frac{1}{(2\pi)^3}\sum_\lambda
   2E_\lambda N_\lambda \psi_\lambda^*
   ({\mathbf p})\psi_\lambda({\mathbf p})\, .
   \label{distP1}
\ee 
Similarly, and under the assumption of a complete factorization of
the two-particle density matrix, the two pion momentum distribution can
be written as
\be
   P_2({\mathbf p_1},{\mathbf p_2})&\equiv&\frac{d^6N}{d^3p_1d^3p_2}
   \nonumber\\
   &=&P_1({\mathbf p_1})P_1({\mathbf p_2})\nonumber\\
   &+&\left|\frac{1}{(2\pi)^3}\sum_\lambda
   2E_\lambda N_\lambda \psi_\lambda^*
   ({\mathbf p_1})\psi_\lambda({\mathbf p_2})\right|^2\, ,
   \label{distP2}
\ee
from where the two-pion correlation function $C_2$ can be written, 
in terms of $P_1$ and $P_2$, as  
\be
   C_2({\mathbf p_1},{\mathbf p_2})\!&=&\!
   \frac{P_2({\mathbf p_1},{\mathbf p_2})}
   {P_1({\mathbf p_1})P_1({\mathbf p_2})}\nonumber\\
   \!&=&\!1\!+\!\frac{\Big|{\displaystyle\sum_{\lambda}}
   E_\lambda N_\lambda \psi_\lambda^*
   ({\mathbf p_1})\psi_\lambda({\mathbf p_2})\Big|^2}
   {{\displaystyle\sum_{\lambda}} E_\lambda N_\lambda 
   \left|\psi_\lambda({\mathbf p_1})\right|^2
   {\displaystyle\sum_{\lambda}}E_\lambda 
   N_\lambda \left|\psi_\lambda({\mathbf p_2})\right|^2}.\nonumber\\
   \label{correl}
\ee
No\-tice that as a con\-se\-quence of the fac\-tor\-iza\-tion 
assumption, the correlation function is such that 
$C_2({\mathbf p},{\mathbf p})=2$. This is usually referred to as 
the completely chaotic pion production scenario~\cite{Gyulassy}, which
is the situation expected to 
occur in a heavy ion collision, given the considerable rescattering
experienced by pions in the production region. In contrast, if the
particles were produced completely coherently, they would occupy a
pure quantum state and the two-pion momentum distribution would 
be simply the product of two single-pion momentum distributions,
leading to the absence of the HBT effect.  

%%%%%%%%%%%%%%%%%%%%%%%%%%%%%%
\begin{figure}[h!] % fig 2
\centerline{\epsfig{file=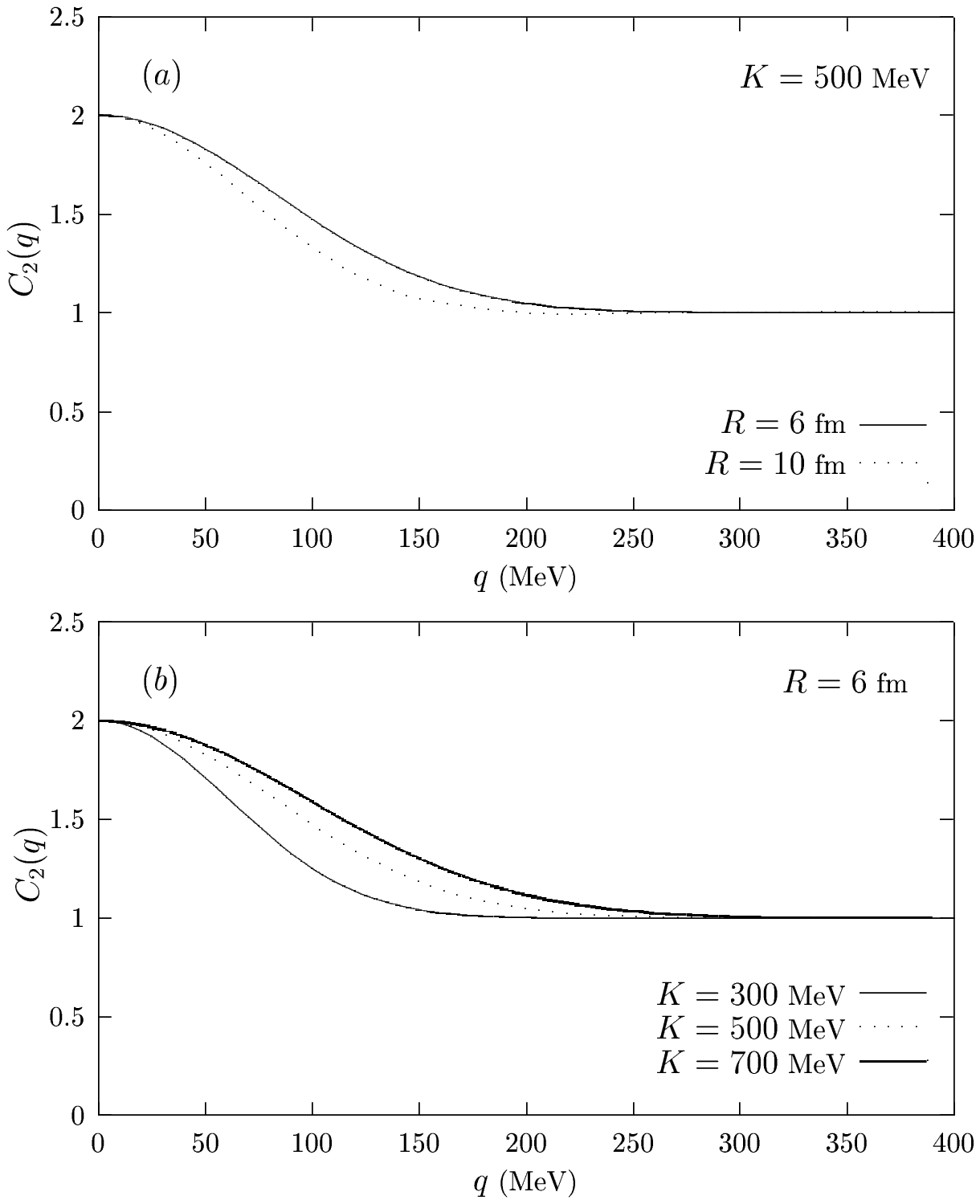,height=4.0in,width=3in}}
\end{figure}
\vspace{-0.8cm}
\begin{figure}[h!] % fig2c
\centerline{\epsfig{file=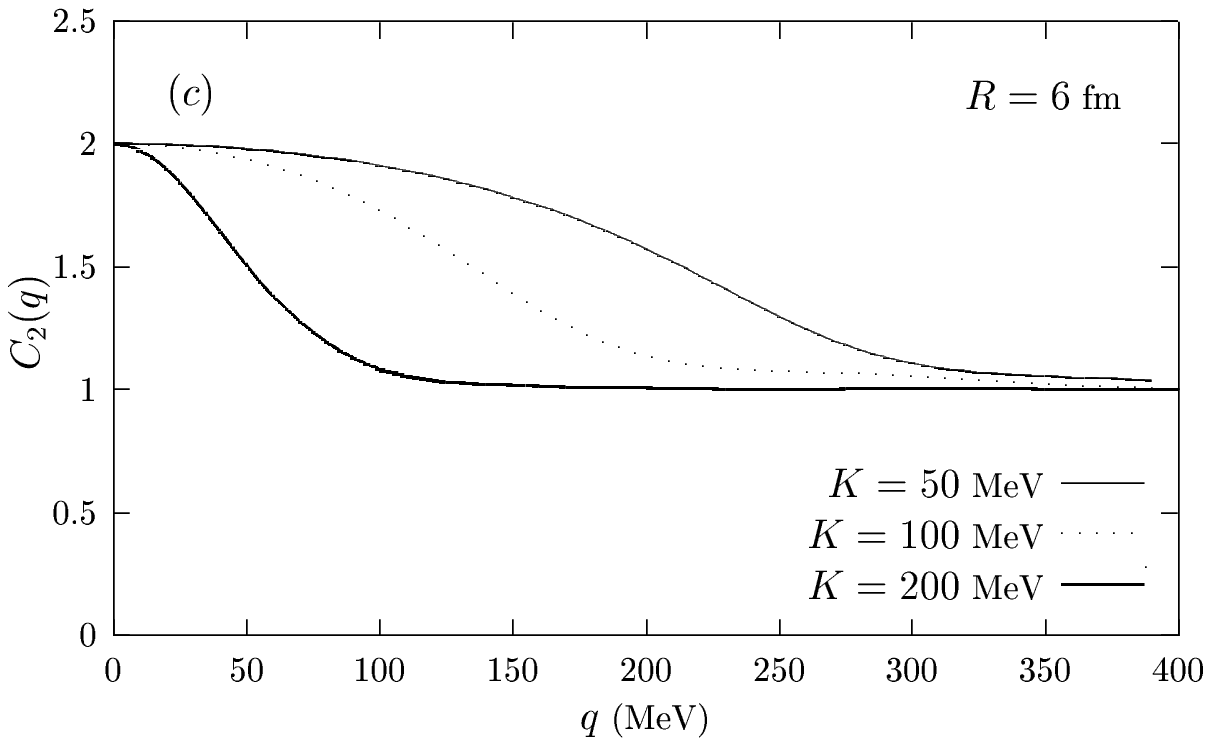,height=2.0in,width=3in}}
\vspace{0.5cm}
\caption{$(a)$: $C_2(q)$ for a fixed value of $K=500$ MeV and for $R=6$ fm
         (solid line) and $R=10$ fm (dotted line). The width of $C_2$
         decreases as $R$ is increased. $(b)$: $C_2(q)$ for a fixed
         value of $R=6$ fm and for $K=300$ MeV (solid line), $K=500$
         MeV (dotted line) and $K=700$ MeV (thick solid line). For the
         chosen values of $K$ the width of $C_2$ increases as $K$ is
         increased. $(c)$: $C_2(q)$ for a fixed
         value of $R=6$ fm and for $K=50$ MeV (solid line), $K=100$
         MeV (dotted line) and $K=200$ MeV (thick solid line). For the
         chosen values of $K$ the width of $C_2$ decreases as $K$ is
         increased. In all cases, the temperature, surface expansion
         velocity and chemical potential have been held fixed to
         $T=120$ MeV, $\beta=0.5 c$ and $\mu=0$ respectively.}
\end{figure}
%%%%%%%%%%%%%%%%%%%%%%%%%%%%%

\section{The effective radius}\label{IV}

Armed with the eigenfunctions describing the confined and expanding
pion system, Eq.~(\ref{solnew}) and with the explicit expression for
the two-pion correlation function in Eq.~(\ref{correl}), it is
possible to perform an analysis to describe the behavior of $C_2$ as a
function of the several variables and parameters involved. For the
spherically symmetric problem described here, the correlation function
depends on the angle between the pion pair momenta. For the sake of
simplicity, let us consider the case in which both momenta ${\mathbf
p_1}$ and ${\mathbf p_2}$ are parallel. In this case, the summation in
the numerator of the second term in Eq.~(\ref{correl}) can be
simplified, with the aid of the addition theorem for the spherical
harmonics
\be
   \sum_{m'=-l}^lY_l^{m'}({\mathbf\hat{p}_1})Y_l^{m'*}({\mathbf\hat{p}_2})
   =\frac{2l+1}{4\pi}\, .
   \label{addi}
\ee
Figure~2 shows the behavior of $C_2(q)$ as a function of $q$, the magnitude
of the pair momentum difference ${\mathbf q}={\mathbf p_2}-{\mathbf
p_1}$. Figure~2$a$ shows $C_2(q)$ for a fixed value of the magnitude
$K=500$ MeV of the average pair momentum ${\mathbf K}=({\mathbf K_2} +
{\mathbf K_1})/2$ for two values of the system radius
$R=6$, $10$ fm. Notice that for a
fixed $K$ the width of the correlation function decreases as $R$ is
increased. Figure~2$b$ shows $C_2(q)$ for a fixed value of $R=6$
fm and three values of $K=300$, $500$, $700$ MeV. For the chosen values of
$K$, the width of the correlation function increases as $K$ is
increased. Figure~2$c$ shows $C_2(q)$ for a fixed value of $R=6$
fm and three values of $K=50$, $100$, $200$ MeV. Notice that in this
case, the width of the correlation function {\it decreases} as $K$ is
increased. In all these figures, the temperature and surface expansion
velocity have been held fixed to $T=120$ MeV and 
$\beta=0.5 c$. These values for $T$ and $\beta$ are chosen in accordance to
the analysis in Ref.~\cite{Esumi} where a correlation between the
transverse flow velocity and the freeze out temperature is found in
such a way that higher temperatures imply lower expansion velocities
and vice versa. Since, at least for AGS energies, not
too high temperatures are reached during the collision, the above
value for $T$ implies that for $\beta$, thus, the free parameters for
the model can be taken either as $R$ and $T$ or $R$ and $\beta$. 

It is also worth mentioning that in both of the above
figures, the value of the chemical potential $\mu$ appearing in
Eq.~(\ref{correl}) has been fixed to $\mu=0$. The behavior of $C_2(q)$
for different values of the chemical potential is shown in
Fig.~3. Notice that varying the chemical 
potential up to values below the onset of Bose-Einstein condensation 
(BEC)~\cite{Ayala1} does not introduce changes in the shape of
the correlation function. This can be understood by noticing that even 
when we increase the system's density and thus the value of $\mu$, we
are not introducing any further correlation among the bound but otherwise
non-interacting set of particles. This situation changes when, for a
given temperature and system's size, $\mu$ is beyond the value to
allow for the ground state to accommodate a significant fraction of the
particle population~\cite{Ayala1}. This is also shown in Fig.~3. In
this case, the ground state population has to be treated separately
from the pions coming from the excited states, since the former
originates from a pure quantum state, as opposed to the assumption
leading to Eq.~(\ref{distP2}). The weight assigned to the pions coming
from the ground state is equal to the ratio of the ground state population
$N_g$ to the total number of pions of the given species
$N$. Correspondingly, the correlation function $C_2(q)$ becomes
flatter and the intercept with the vertical axis occurs for values smaller
than $2$, that is, $\lambda=1-N_g/N$. Thus, as the density
increases, this behavior signals that for the given temperature and
volume, pions are predominantly emitted from the ground state. This is
in agreement with the analysis in Ref.~\cite{Pratt2}. 

%%%%%%%%%%%%%%%%%%%%%%%%%%%%%%
\begin{figure}[t!] % fig 3
\centerline{\epsfig{file=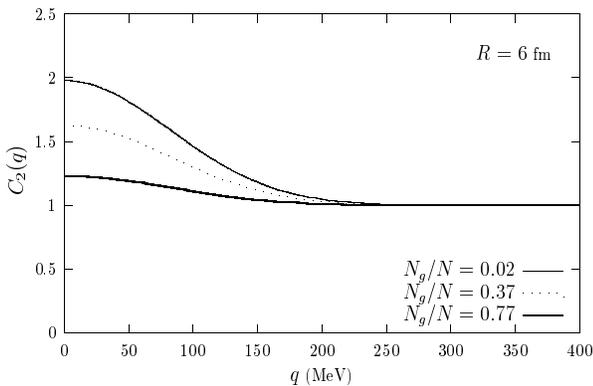,height=2.0in,width=3in}
}
\vspace{0.5cm}
\caption{$C_2(q)$ for a fixed value $R=6$ fm and different values of
         the ground state population fraction. For $\mu=0$ (solid line) the
         ground state population is negligible and
         $\lambda\sim 1$. However, when the ground state population
         becomes a significant fraction of the total multiplicity, 
         {\it e.g.} for $\mu=315$ MeV, $N_g/N=0.37$ and 
         $\lambda\sim 0.62$ (dotted line). When almost all of the
         particles occupy the ground state, {\it e.g.} for $\mu=317.5$
         MeV, $N_g/N=0.77$, $\lambda\sim 0.23$ (thick
         solid line) and the correlation function becomes flatter.}
\end{figure}
%%%%%%%%%%%%%%%%%%%%%%%%%%%%%

Another property of the bound and expanding system of pions that can be
extracted from the correlation function $C_2(q)$ is the behavior of
the system's effective radius $R_{\mbox{\small{eff}}}$ as a function of
$K$. The relevant quantity to pay attention to is the ratio 
$\eta=T/\gamma (R)m\beta$ of the energy scale associated with random motion, 
{\it i.e.} $T$, to the energy scale associated with ordered motion, {\it i.e.} 
$\gamma (R)m\beta$. For $K$ small compared to $\eta T$ --that is, when
the average pair momentum is mostly due to random motion-- 
$R_{\mbox{\small{eff}}}$ is an increasing function of $K$. This can be
understood by noticing~\cite{Zhang} that increasing $K$ corresponds to
increasing both of the momenta in the pion pair. Correspondingly, the
quantum states that contribute to the momentum distributions $P_1$ and
$P_2$ are those with increasingly larger quantum numbers. But,
according to Eq.~(\ref{occupation}), these states are suppressed by
their statistical weight and therefore, only those other states with
smaller quantum numbers --and consequently with a larger spread in 
coordinate space-- can contribute significantly to the correlation
function, which in turn drops faster as a function of $q$ with
increasing $K$, leading to an increase in size of the apparent region of
particle emission. However, for $K > \eta T$, the collective motion
dominates over the thermal component in 
$K$ and the relevant physical effect that dictates the behavior of 
$R_{\mbox{\small{eff}}}$ is the correlation between the spatial region
of emission of pions and the pair momentum introduced by the
collective expansion, in such a way that faster pions are more likely
to be emitted from points close in space~\cite{Pratt} leading to a
reduction in size of the apparent region of particle emission. 

%%%%%%%%%%%%%%%%%%%%%%%%%%%%%%
\begin{figure}[t!] % fig 4
\centerline{\epsfig{file=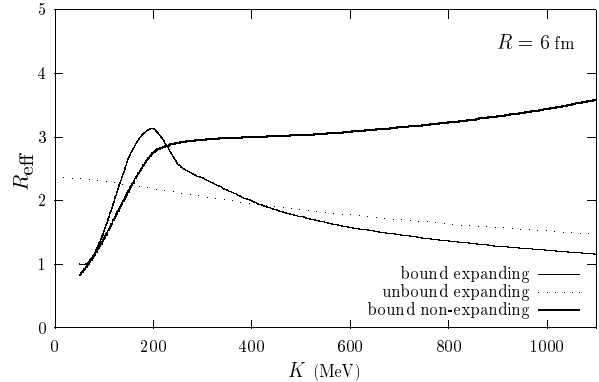,height=2.0in,width=3in}}
\vspace{0.5cm}
\caption{$R_{\mbox{\small{eff}}}$ for a fixed value $R=6$ fm as a
         function of $K$ for an expanding and bound system (solid
         line), an expanding and unbound system (dotted line) and a
         bound and non-expanding system (thick solid line). Notice that
         in the first case, $R_{\mbox{\small{eff}}}$ reaches a maximum
         at a value of $K\sim\eta T$, whereas for the second case,
         $R_{\mbox{\small{eff}}}$ decreases with increasing $K$ and in the
         third case $R_{\mbox{\small{eff}}}$ grows with increasing
         $K$. Also, for the expanding and unbound system, the
         parameter $R_{\mbox{\tiny{gauss}}}$ has been chosen in such a
         way as to give the same r.m.s radius than a rigid sphere with
         $R=6$ fm.}
\end{figure}
%%%%%%%%%%%%%%%%%%%%%%%%%%%%%

The behavior of $R_{\mbox{\small{eff}}}$ as a function of $K$ is shown
in Fig.~4 (solid line). Notice that the curve shows a maximum for a
value of $K\sim\eta T$. The curve is obtained by fitting
the correlation functions $C_2(q)$ to Gaussians of the form
\be
   g(q)=1+\exp (-q^2R^2_{\mbox{\small{eff}}})\, .
   \label{gaus}
\ee
Equation~(\ref{gaus}) is a good description for correlation functions
with large $K$. For small values of $K$, the fit is not as good. For
comparison, also shown in Fig.~4 is the behavior of
$R_{\mbox{\small{eff}}}$ for an expanding system without a boundary
(dotted line) and for a bound system without the effects
hydrodynamical expansion (thick solid line). For the former we choose
a spherically symmetric phase space Gaussian distribution given by 
\be
   G({\mathbf x},{\mathbf p})=e^{-{\mathbf x}^2/2R_{\mbox{\tiny{gauss}}}^2}
   e^{-\gamma (x)\left(E_p-{\mathbf v\cdot p}\right)/T}\, ,
   \label{gausin}
\ee
with $E_p=\sqrt{{\mathbf p}^2+m^2}$ and ${\mathbf v}$ and $\gamma (x)$
given by Eqs.~(\ref{paramvel}) and~(\ref{gammaapprox}), respectively.
The correlation function is given in terms of 
$G({\mathbf x},{\mathbf p})$ by~\cite{Pratt}
\be
   C_2({\mathbf p_1},{\mathbf p_2})\!=\!1\!+\!\frac{\left| \int d^3x\,
   G({\mathbf x},\frac{{\mathbf p_1}+{\mathbf p_2}}{2})
   e^{-i({\mathbf p_1}-{\mathbf p_2})\cdot{\mathbf x}}\right|^2}
   {\left(\int d^3x\,G({\mathbf x},{\mathbf p_1})\right)
   \left(\int d^3x\,G({\mathbf x},{\mathbf p_2})\right)}.
   \label{correl2}
\ee
For the bound and non-expanding system, the eigenfunctions are given
in terms of Bessel functions of the first kind~\cite{Ayala1} (see also
Ref.~\cite{Zhang}). The corresponding expression for
$R_{\mbox{\small{eff}}}$ is obtained from that of $C_2(q)$ by also
fitting Gaussians of the form given by Eq.~(\ref{gaus}). Notice that
the curve representing the effective radius as a function of $K$ for 
a bound but non-expanding system grows with $K$, in agreement with the
analysis of Ref.~\cite{Zhang}. In contrast, the curve representing the
effective radius for an unbound but expanding system decreases
monotonically as $K$ is increased, also in agreement with the analysis of 
Ref.~\cite{Pratt}.

\section{Conclusions}\label{V}
In this work, we have studied the effects that a confining boundary
together with hydrodynamical expansion at freeze out, play on the
two-pion correlation function, in the context of relativistic
heavy-ion collisions. We have argued that the confining boundary could
be produced as a consequence of the high pion density that can be
achieved at freeze out in central collisions. 

We have shown that for a given system's volume and temperature, 
varying the multiplicity, and therefore the chemical potential, does not
introduce any changes in the correlation function when $\mu$ is below
the values for BEC. However, the intercept of the function $C_2(q)$
occurs for values less than $2$ when the chemical potential is beyond
the value to allow for BEC. A similar behavior can be expected for a
given pion density if the freeze out temperature is below the critical
temperature for BEC. However, this is a less likely scenario in this
kind of collisions. 

We have found the behavior of $C_2(q)$ when varying either $R$ or $K$
keeping the other variable fixed. Since the importance of correlation
analyses rests basically on the information that it can provide about
the physical size of the system produced during the collision, a main
result of the present work is the functional dependence of the
effective system radius $R_{\mbox{\small{eff}}}$ with the magnitude of
the average pair momentum $K$. We have shown
that the interplay of the energy scales associated with collective and
random motion, $\gamma (R)m\beta$ and $T$, respectively, produce a
maximum for $R_{\mbox{\small{eff}}}$ at a value $K\sim\eta T$, where
$\eta=T/\gamma (R)m\beta$. The physical origins of this behavior are
the combined effects of the confining boundary and hydrodynamical
expansion. In the regime where $K$ is basically due to random motion,
the boundary effects are the most important and
$R_{\mbox{\small{eff}}}$ grows as a function of $K$. However, in the
regime where $K$ is basically due to collective expansion, the
effective size of the system is dictated by the correlation between
the points of emission and the pair momentum and
$R_{\mbox{\small{eff}}}$ decreases as a function of $K$.

\section*{Acknowledgments}
Support for this work has been received in part by CONACyT-M\'exico
under grants Nos. 29273-E and 32279-E.

\end{document}